\documentclass[11pt, reqno,preprint]{article}
\usepackage{jheppub}
\usepackage{epsfig}
\usepackage{amssymb}
\usepackage{amsmath}
\usepackage{mathrsfs}
\usepackage{hyperref}
\usepackage{multirow}

\def\be{\begin{equation}}
\def\ee{\end{equation}}
\def\ba{\begin{eqnarray}}
\def\ea{\end{eqnarray}}
\def\nl{\nonumber\\}

\def\a{\alpha}

\def\ad{{\dot\alpha}}
\def\bd{{\dot\beta}}

\def\lambdab{\tilde\lambda}
\def\Lambdab{\tilde\Lambda}
\def\etab{\tilde\eta}
\def\mub{\tilde\mu}
\def\cb{\tilde c}
\def\Cb{\tilde C}

\def\ZZ{\mathcal{Z}}
\def\WW{\mathcal{W}}
\def\MM{\mathcal{M}}

\def\l{\langle}
\def\r{\rangle}

\def\vGL{\textrm{vol\,GL}}

\title{A Link Representation for Gravity Amplitudes}

\author{Song He}

\affiliation{Max-Planck-Institut f\"ur Gravitationsphysik, Am M\"uhlenberg 1, 14476 Potsdam, Germany}

\emailAdd{songhe@aei.mpg.de}

\abstract{We derive a link representation for all tree amplitudes in $\mathcal{N}=8$ supergravity, from a recent conjecture by Cachazo and
Skinner. The new formula explicitly writes amplitudes as contour integrals over constrained link variables, with an integrand naturally
expressed in terms of determinants, or equivalently tree diagrams. Important symmetries of the amplitude, such as supersymmetry, parity and
(partial) permutation invariance, are kept manifest in the formulation. We also comment on rewriting the formula in a GL($k$)-invariant manner,
which may serve as a starting point for the generalization to possible Grassmannian contour integrals.}

\begin{document}

\maketitle

\section{Gravity amplitudes from rational curves}

A new formulation of the classical S-matrix in $\mathcal{N}=8$ supergravity has been conjectured recently~\cite{Cachazo:2012kg}, which expresses
$n$-particle, N$^{k{-}2}$MHV amplitudes in terms of degree $k{-}1$ holomorphic map to twistor space~\footnote{The formula, together with a
related but distinct ``twistor-string'' inspired formula~\cite{Cachazo:2012da}, are partly based on a new expression for the MHV gravity
amplitudes~\cite{Hodges:2012ym}.},

\be \MM_{n,k}(\lambda_a,\lambdab_a,\etab_a)=\int\frac{\prod^{k{-}1}_{d=1} d^{4|8}\ZZ'_d}{\vGL (2)}\,{\det}'(\Phi)\,{\det}'(
\tilde{\Phi})\,\prod^n_{a=1} d^2\sigma_a
\delta^2(\lambda_a-\lambda(\sigma_a))\exp[-i\,\mu(\sigma_a)\cdot\lambdab_a+i\,\eta(\sigma_a)\cdot\etab_a]~,\ee where the map from the worldsheet
to $\mathcal{N}=8$ supertwistor space, $\ZZ=(\lambda^{\a},\mu^{\ad},\eta^A)$, is given by a degree $k{-}1$ polynomial of worldsheet coordinates
$\sigma^{\underline{\alpha}}=(\sigma^{\underline{1}},\sigma^{\underline{2}})$, \be \ZZ(\sigma)=\sum^{k{-}1}_{d=0} \ZZ'_d
(\sigma^{\underline{1}})^d(\sigma^{\underline{2}})^{k{-}d{-}1}~.~\label{twistorstring}\ee Up to an overall GL(2) transformation, one integrates
over $\prod^{k{-}1}_{d=0}d^{4|8}\ZZ'_d$, and over the vertex operator insertions, $\prod^n_{a=1} d^2\sigma_a$. In the integrand, there are $n$
external twsitor-space wavefunctions for momentum eigenstates $(\lambda^{\a},\lambdab_{\ad},\etab_A)_a$, $a=1,...,n$. Note $\a=1,2$,
$\ad=\dot{1},\dot{2}$ are SU(2) spinor indices, $A=1,...,8$ is the SU(8) R-symmetry index, and the inner products are defined as
$\mu\cdot\lambdab:=\mu^{\ad}\lambdab_{\ad}$, $\eta\cdot\etab:=\eta^A\etab_A$~.

Remarkably, Cachazo and Skinner were able to observe that the non-trivial content of all gravity tree amplitudes is hidden in the two
``determinant'' factors in~(\ref{twistorstring}), which can be obtained from minors of two $n\times n$ matrices, $\tilde{\Phi}$ and
$\Phi$~\cite{Cachazo:2012kg},

\ba \tilde{\Phi}_{a\,b, a\neq b}&=\displaystyle \frac{[a\,b]}{(a\,b)}~, \quad \tilde{\Phi}_{a\,a}&=-\sum_{b\neq a} \tilde{\Phi}_{a\,b}
\frac{\tilde{y}_b}{\tilde{y}_a}~,\nl \Phi_{a\,b, a\neq b}&=\displaystyle \frac{\l a\,b \r}{(a\,b)}~, \quad \Phi_{a\,a}&=-\sum_{b\neq a}
\Phi_{a\,b}\frac{\prod_{c\neq a} (a\,c)}{\prod_{d\neq b} (b\,d)} \frac{y_b}{y_a}~\label{Phi},\ea which depend on brackets of external spinors,
$\l a\, b\r:=\epsilon_{\a\beta}\lambda^{\a}_a\lambda^{\beta}_b$, $[a\,b]:=\epsilon^{\ad \bd}\lambdab_{a \ad}\lambdab_{b \bd}$, on those of
worldsheet insertions $(a\,b):=\epsilon_{\underline{\a}\underline{\beta}}\sigma_a^{\underline{\a}}\sigma_a^{\underline{\beta}}$, and on
worldsheet reference points $p_1,...,p_k$, $q_1,...,q_{n{-}k}$, through the combinations, \be\tilde{y}_a:=\prod^k_{i=1} (a\,p_i)~,\quad
y_a:=\prod^{n{-}k}_{i=1} (a\,q_i)~.\ee

Note that $\tilde{\Phi}$ has rank $n{-}k{-}1$, and $\Phi$ has rank $k{-}1$, thus we need to define minors $|\tilde{\Phi}_{\textrm{red}}|$ by
deleting $k{+}1$ rows and $k{+}1$ columns of $\tilde{\Phi}$, similarly minors $|\Phi_{\textrm{red}}|$ by deleting $n{-}k{+}1$ rows and
$n{-}k{+}1$ columns of $\Phi$. As shown in~\cite{Cachazo:2012kg}, the ``determinants'' needed in~(\ref{twistorstring}) are the following ratios,
which are independent of what rows and columns we choose to delete, and are symmetric under any permutations of $n$ particles,

\be {\det}'(\tilde{\Phi}):=\frac{|\tilde{\Phi}_{\textrm{red}}|}{|\tilde{r}_1\cdots\tilde{r}_{k{+}1}||\tilde{c}_1\cdots\tilde{c}_{k{+}1}|}~,
\quad {\det}'(\Phi):=\frac{|\Phi_{\textrm{red}}|}{|r_1\cdots r_{k{-}1}||c_1\cdots c_{k{-}1}|}~\label{determinants},\ee where in the
denominators, we have Vandermonde determinants, obtained from all rows that are either removed or remained, \be
|\tilde{r}_1\cdots\tilde{r}_{k{+}1}|=\prod_{a<b\,\in \{\textrm{removed}\}}(a\,b)~, \quad |r_1\cdots r_{k{-}1}|=\prod_{a<b\,\in
\{\textrm{remain}\}}(a\,b)~,\ee and similarly Vandermonde determinants of removed or remained columns.

Except for the ``determinants'', the formula~(\ref{twistorstring}) resembles the connected prescription of Witten's twistor string
theory~\cite{Witten:2003nn}~\cite{Roiban:2004yf}, for tree amplitudes in $\mathcal{N}=4$ super--Yang-Mills (SYM). However, as is well known, it
is highly non-trivial to carry out the integration over the worldsheet insertions, since it involves solving polynomial equations. To overcome
the difficulties, it has proved very useful to rewrite the connected prescription into the so-called link
representation~\cite{Spradlin:2009qr}~\cite{Dolan:2009wf}, first introduced in~\cite{ArkaniHamed:2009si}.

This has enabled the computation of all tree amplitudes in SYM using the connected prescription~\cite{Dolan:2010xv}~\cite{Dolan:2011za}, and has
led to a remarkable duality~\cite{Nandan:2009cc}~\cite{ArkaniHamed:2009dg}, linking the twistor-string prescription to the tree-contour of the
Grassmannian integral~\cite{Grassmannian}, which has been shown to encode leading singularities of all-loop amplitudes in SYM. In this note we
derive a link representation for gravity amplitudes, from this ``twistor-string''-like formula (\ref{twistorstring}). We hope the new
formulation will not only improve our understanding of tree-level gravity amplitudes, but also shed light on a possible Grassmannian integral
for leading singularities in $\mathcal{N}=8$ supergravity.

\section{Gravity amplitudes in a link representation}

In~\cite{Feng:2012sy}, Feng and the author introduced a rewriting of the determinant formula of tree-level MHV gravity
amplitudes~\cite{Hodges:2012ym}. Here, following similar ideas, we find it useful to define a more symmetric version of the matrices defined
in~(\ref{Phi}), \be\tilde{\Psi}_{a b}=\tilde{\Phi}_{a b}\,\tilde{y}_a\,\tilde{y}_b~,\quad \Psi_{a b}=\Phi_{a b}\,\frac{y_a\,y_b}{\prod_{c\neq a}
(a\,c)\,\prod_{d\neq b}(b\,d)}~\label{Psi},\ee which have a nice property that each row or column sums up to zero. Furthermore, we divide the
$n$ indices, $a=1,...,n$, into two sets, $N$ and $P$, of $k$ and $n{-}k$ indices, respectively, e.g. $N=\{1,...,k\}$ and $P=\{k{+}1,...,n\}$.
Then it is convenient to make the following special choice for reference points: we set $p$'s to coincide with worldsheet insertions, $\sigma_I$
for $I\in N$, and $q$'s to coincide with $\sigma_i$ for $i\in P$~\footnote{In the following we will always use $I,J,...$ for labels in $N$,
$i,j,...$ for labels in $P$, and $a,b,...$ for those in $N\cup P$, and will often abbreviate the sum (product) over $I\in N$, or $i\in P$,
simply as $\sum_I$ ($\prod_I$), or $\sum_i$ ($\prod_I$).}. With this choice, $\tilde{y}_I=0$ and $y_i=0$, which immediately kills all entries of
the form $\tilde{\Psi}_{i b}$, $\tilde{\Psi}_{a j}$, and those of the form $\Psi_{I b}$, $\Psi_{a J}$.

This simplifies the formula significantly, since $\tilde{\Psi}$ and $\Psi$ are effectively reduced to $(n{-}k)\times (n{-}k)$ and $k\times k$
matrices, respectively, \be \tilde{\Psi}_{i j}=\tilde{\Phi}_{i j}\,x_i\,x_j~,\quad \Psi_{I J}=\frac{\Phi_{I J}}{x_I\,x_J}~\label{psi},\ee where,
by~(\ref{Psi}), we have introduced variables $x_a$ to relate the matrices to $\tilde{\Phi}_{i j}$ and $\Phi_{I J}$, \be x_a:=\prod_{J\neq
a,\,J\in N}(a\,J)~.\ee Thus, up to an overall factor of $x_a$'s, we can compute the minor $|\Phi_{\textrm{red}}|$
($|\tilde{\Phi}_{\textrm{red}}|$) by deleting one row and one column of $\Psi_{I J}$ ($\tilde{\Psi}_{i j}$), which we will write down explicitly
in a moment.

As a consequence of the reduction of matrices, ${\det}'(\Phi)$ now only depends on external twistors $\ZZ_I$ for $I\in N$, and
${\det}'(\tilde{\Phi})$ only depends on external dual twistors $\WW_i$ for $i\in P$, where $\WW=(\mub_{\a},\lambdab_{\ad},\etab_A)$~\footnote{
Note $\l I\,J\r=\ZZ_I \mathbf{I} \ZZ_J$ and $[i\,j]=\WW_i \mathbf{I} \WW_j$, where the infinity twistor, $\mathbf{I}$, projects any twistor
$\ZZ$ (dual twistor $\WW$) to its $\lambda$ ($\lambdab$) component.}. It will be useful to define inner products of twistors and dual twistors,
$\WW\cdot \ZZ=\mub\cdot\lambda-\mu\cdot\lambdab+\eta\cdot\etab$, with $\mub\cdot\lambda:=\mub_{\a}\lambda^{\a}$.

We have seen that in this ``gauge'' (choice of reference points, as well as that of deleted rows and columns), the formula wanted to be
transformed into the basis of $\{\ZZ_I, \WW_i\}$, \be \MM(\ZZ_I, \WW_i)=\prod_i\int d^2\lambda_i\,e^{i\mub_i\cdot\lambda_{i}} \prod_I \int
d^2\lambdab_I\,d^{0|4}\etab_I\,e^{i (\mu_I\cdot\lambdab_I-\eta_I\cdot\etab_I)}\,\MM(\lambda_a,\lambdab_a,\etab_a),\ee which is exactly the
natural basis for defining a link representation. Under the transformation, the ``determinants'' stay unchanged, and the product of
wavefunctions becomes

\be \prod_i \exp[ i\,\WW_i \cdot \ZZ(\sigma_i)]\prod_I \delta^{4|8}(\ZZ_I-\ZZ(\sigma_I))~.\ee

At this point, it is clear that we can trivially use the $4k$ bosonic functions to carry out the integration over $\ZZ'$, which sets the
polynomial to be \be \ZZ(\sigma)=\sum_I \ZZ_I \prod_{J\neq I}\frac{(\sigma\,J)}{(I\,J)}~.\ee Unlike the SYM case, the integration gives a
Vandermonde Jacobian $\prod_{I<J}(I\,J)^4=\prod_I x^2_I$~, thus

\be \MM_{n,k}(\ZZ_I, \WW_i)=\prod_I x^2_I\int \frac{d^{2n} \sigma_a}{\vGL(2)} {\det}'(\Phi) {\det}'(\tilde{\Phi}) \exp[ i \sum_{i,I} c_{I
i}\,\WW_i \cdot \ZZ_I]~,\label{twistor}\ee where we have defined $k\times(n{-}k)$ link variables ``linking'' $\ZZ_I$'s and $\WW_i$'s, \be c_{I
i}=\prod_{J\neq I}\frac{(i\, J)}{(I\,J)}=\frac{x_i}{x_I (i\, I)}~.\label{link}\ee In this form it is straightforward to transform back to
momentum space~\cite{ArkaniHamed:2009si}, \be \mathcal{M}_{n,k}(\lambda_a,\lambdab_a,\etab_a)=\prod_I x^2_I \int \frac{d^{2n} \sigma_a}{\vGL(2)}
{\det}'(\Phi) {\det}'(\tilde{\Phi})\delta^2(\lambda_i-c_{I i}\lambda_I)\delta^{2|8}(\Lambdab_I+c_{I i}\Lambdab_i)~\label{momspace}, \ee where
$\Lambdab_a:=(\lambdab_a,\etab_a)$, and the products over $i,I$ and sums over $I,i$ are kept implicit.

Now we want to explicitly simplify the ``determinants'' in this gauge. Using a particular way of expanding such determinants, the so-called
Matrix-Tree Theorem (for details of the theorem and applications, see~\cite{Feng:2012sy}), we note that, up to overall factors,
$|\Phi_{\textrm{red}}|$, $|\tilde{\Phi}_{\textrm{red}}|$ can be expanded as summing over all trees with weighted edges of the form~(\ref{psi}),
and vertices in $N$ and $P$ respectively. Collecting the denominators, the result is very simple,

 \ba {\det}'(\Phi)&=&(-)^{k{-}1}\prod_I x_I \sum_{T^-\in
\mathcal{T}(N)}\prod_{\{I\,J\}\in E(T^-)}\frac{\l I\,J\r}{(I\,J) x_I x_J}~,\nl
    {\det}'(\tilde{\Phi})&=&(-)^{n{-}k{-}1}\prod_I\frac 1{x_I}\prod_i\frac 1{x_i^2}\sum_{T^+\in \mathcal{T}(P)}\prod_{\{i\,j\}\in E(T^+)}\frac{[i\,j]x_i
    x_j}{(i\,j)}~,
\ea where $\mathcal{T}(V)$ is the set of all trees with vertices in $V$, and $E(T)$ contains all edges $\{i\,j\}$, labeled by endpoints $i,j$,
in a tree $T$. The full integrand, $F_{n,k}:=\prod_I x^2_I\,{\det}'(\Phi){\det}'(\tilde{\Phi})$, looks quite symmetric between $N$ and $P$
(which is related to parity invariance, as we will see soon),

\be F_{n,k}=(-)^n\prod_I x^2_I\sum_{T^-\in \mathcal{T}(N)}\,\prod_{\{I\,J\}\in E(T^-)}\frac{\l I\,J\r}{(I\,J)x_I x_J}\prod_i \frac1{x_i^2}
\,\sum_{T^+\in \mathcal{T}(P)}\,\prod_{\{i\,j\}\in E(T^+)}\frac{[i\,j] x_i x_j}{(i\,j)}~.\label{F} \ee

The next step is to rewrite everything in terms of link variables $c_{I i}$, and the computation is similar to that of~\cite{Dolan:2009wf}. We
formally insert an identity, \be 1=\int d^{k\times(n{-}k)} c_{I i}\prod_{I\,i}\delta(c_{I i}-\prod_{J\neq I}\frac{(i\, J)}{(I\,J)})~,\ee which
fixes the link variables as~(\ref{link}), and enables us to integrate over worldsheet insertions using $2\,n{-}4$ out of the $k\times(n{-}k)$
delta functions. A convenient choice is that~\cite{Dolan:2009wf}, given any $R,S\in N$, $r,s\in P$ (we denote $N':=N/\{R,S\}$, $P':=P/\{r,s\}$),
we use delta functions of $c'\in \{c_{R i}, c_{S i}, c_{I r}, c_{I s}: i\in N', I\in P\}$ to integrate over worldsheet insertions. We can also
fix the GL(2) gauge by singling out $\sigma_R,\sigma_S$, \be \frac 1{\vGL(2)}\,\int \prod^n_{a=1} d^{2\,n}\sigma_a =(R\,S)^2\,\int \prod_{a\neq
R,S} d^{2\,n{-}4}\sigma_a ~.\ee The integration over $\sigma$'s then gives a Jacobian $J_1$,

\be J_1=\prod_i \left|\frac{\partial (c_{R i}, c_{S i})}{(\sigma^{\underline{1}}_i,\sigma^{\underline{2}}_i)}\right|^{-1} \prod_{I\in N'}
\left|\frac{\partial (c_{I r}, c_{I s})}{(\sigma^{\underline{1}}_I,\sigma^{\underline{2}}_I)}\right|^{-1}=\prod_i \frac{x_i^2}{ c^2_{R
i}\,c^2_{S i}\,x_R\,x_S\,(R\,S)} \prod_{I\in N'} \frac{x_r\,x_s}{x_I^2\,c^2_{I r}\,c^2_{I s}\,(r\,s)}~.\label{J1}\ee

Relations~(\ref{link}) are over-constrained, and imply $m:=(k{-}2)\times(n{-}k{-}2)$ constraints on the link variables, $c_{I i}$. As shown
in~\cite{Dolan:2009wf}, the remaining $m$ delta functions, labeled by $\cb \in \{c_{I i}: I\in N', i\in P'\}$, are equivalent to $m$ sextics. In
the choice above, we write $\Cb_{I i}:=C_{R S I\,r s i}$, which are given by the determinant of a $3\times 3$ matrix, \be
\Cb_{I i}=\left|\begin{array}{ccc} c_{R s}\,c_{R i} & c_{R i}\,c_{R r} & c_{R r}\,c_{R s}\\
c_{Ss}\,c_{Si} & c_{Si}\,c_{Sr} & c_{Sr}\,c_{Ss}\\
c_{Is}\,c_{Ii} & c_{Ii}\,c_{Ir} & c_{Ir}\,c_{Is}\\
\end{array} \right|~,\ee and explicitly, in terms of a basic object, $c^{I J}_{i\,j}=c_{I i}\,c_{J j}-c_{I j}\,c_{J i}$, the sextics are given by, \be
\Cb_{I\,i}=c_{I r}\,c_{S i}\,c^{R S}_{r\,s}\,c^{R I}_{s\,i}-c_{I i}\,c_{S r}\,c^{R S}_{s\,i}\,c^{R I}_{r\,s}~.\label{sextics} \ee The
transformation from $\delta(\cb)$'s to $\delta(\Cb)$'s results in another Jacobian~\cite{Dolan:2009wf},

\be J_2=\prod_{I\in N', i\in P'} \frac{\partial \Cb_{I\,i}}{\partial \cb_{I\,i}}=(c_{R\,S\,;\,r\,s})^m\,\prod_{I\in
N'}(c_{I\,r}\,c_{I\,s})^{n{-}k{-}2}\,\prod_{i\in P'}(c_{R\,i}\,c_{S\,i})^{k{-}2}\,\prod_{I\in N', i\in P'}\frac 1{c_{I\,i}}~.\label{J2}\ee

Now we can express the amplitude as an integral over $k\times(n{-}k)$ link variables, \be \MM_{n,k}(\ZZ_I, \WW_i)=\int d^{k\times(n{-}k)} c_{I
i}\,J_1\,J_2\,(R\,S)^2\,F_{n,k}\,\exp[ i\,c_{I\,i} \WW_i\cdot\ZZ_I]\prod'_{I,i}\,\delta(\Cb_{I\,i})~\label{link1},\ee where we have use
$\prod'_{I,i}$ to denote $\prod_{I\in N', i\in P'}$. Note the following combination can be written in terms of link variables,

\be \frac{x_i\,x_j}{x_I\,x_J\,(I\,J)\,(i\,j)}=\frac{c_{I\,i}c_{J\,j}c_{I\,j}c_{J\,i}}{c^{IJ}_{i\,j}}.\ee Since there are exactly $k{-}1$ edges
in any tree $T^-$ , and $n{-}k{-}1$ ones in any $T^+$, the weights in~(\ref{F}) can always be combined with factors in $J_1$, (\ref{J1}), to
produce such combinations, \ba (R\,S)^2\,J_1\,F_{n,k}&=&(-)^n c^{R S}_{r\,s}\,c_{R r}\,c_{S s}\,c_{R s}\,c_{S r}\,\prod_I\frac 1{c^2_{I
r}\,c^2_{I s}} \prod_i\frac 1{ c^2_{R i}\,c^2_{S i}}\times\nl &&\sum_{T^-}\,\prod_{\{I\,J\}}\frac{\l I\,J \r\,c_{I r}\,c_{J s}\,c_{I s}\,c_{J
r}}{c^{I J}_{r\,s}}\,\sum_{T^+}\,\prod_{\{i\,j\}}\frac{[ i\,j]\,c_{R i}\,c_{S j}\,c_{R j}\,c_{S i}}{c^{R S}_{i\,j}}~.\label{combination}\ea
Since $J_2$ is a function of link variables, we have seen that the formula was written in a link representation, given explicitly by
(\ref{link1}) and (\ref{combination}).

To put it in a more compact form, using the Matrix-Tree Theorem~\cite{Feng:2012sy}, we can write the result in terms of determinants, obtained
after deleting any one row and one column of the following matrices,

\ba \Phi^-_{I J, I\neq J}&=\displaystyle \frac{\l I\,J\r}{c^{I J}_{r\,s}}~,\quad \Phi^-_{I I}&=-\sum_{J\neq I}\Phi^-_{I J}\,\frac{c_{I r}\,c_{I
s}}{c_{J r}\,c_{J s}}~,\nl \Phi^+_{i j, i\neq j}&=\displaystyle \frac{[ i\,j ]}{c^{R S}_{i\,j}}~,\quad \Phi^+_{i i}&=-\sum_{j\neq i}\Phi^+_{i
j}\,\frac{c_{R i}\,c_{S i}}{c_{R j}\,c_{S j}}~. \ea If we delete row $P$ ($p$) and column $Q$ ($q$) from $\Phi^-$ ($\Phi^+$), the desired
expansion is given by \be \frac{|\Phi^-|^P_Q}{c_{P r}\,c_{P s}\,c_{Q r}\,c_{Q s}}\frac{|\Phi^+|^p_q}{c_{R p}\,c_{S p}\,c_{R q}\,c_{S q}}~,\ee
where $|\Phi|^a_b$ means the minor with row $a$ and column $b$ deleted from $\Phi$. Staring at the prefactor in~(\ref{combination}), it is
natural to choose $\{P,Q\}=\{R,S\}$ and $\{p,q\}=\{r,s\}$, which gives,

\be \MM_{n,k}(\ZZ_I, \WW_i)=\int d^{k\times (n{-}k)} c_{I i}\,J\,|\Phi^-|^R_S\,|\Phi^+|^r_s\exp[ i\,c_{I i}
\WW_i\cdot\ZZ_I]\prod'_{I,i}\,\delta(\Cb_{I i})~,\label{linkrep1}\ee and the momentum space formula, \be
\MM_{n,k}(\lambda_a,\lambdab_a,\etab_a)=\int d^{k\times (n{-}k)} c_{I i}\,J\,|\Phi^-|^R_S\,|\Phi^+|^r_s\,\delta^2(\lambda_i-c_{I
i}\lambda_I)\,\delta^{2|8}(\Lambdab_I+c_{I i}\Lambdab_i)\prod'_{I,i}\,\delta(\Cb_{I i})~,\label{linkrep2}\ee where
$J=\frac{c^{R\,S}_{r\,s}}{c_{R\,r}\,c_{S\,s}\,c_{R\,s}\,c_{S\,r}}\,J_2$ with $J_2$ given by~(\ref{J2}), and $\Cb_{I i}$ is given
by~(\ref{sextics}). Thus we have obtained the link representation of gravity tree amplitudes.

In addition to $m$ sextics, we have $2\,n$ bosonic delta functions in~(\ref{linkrep2}), which, after pulling out $4$ of them corresponding to
momentum conservation, can be used to integrate out $2\,n{-}4$ link variables, e.g. $c'\in \{c_{R i}, c_{S i}, c_{I r}, c_{I s}: i\in N', I\in
P\}$. Taking into account the Jacobian~\cite{Dolan:2009wf}, we obtain an equivalent form of the link representation in momentum space, \be
\MM_{n,k}(\lambda_a,\lambdab_a,\etab_a)=\frac{\delta^4(\sum_a \lambda_a\lambdab_a)}{\l R\,S\r^{n{-}k{-}2}[r\,s]^{k{-}2}}\int d^m
\cb\,J\,|\Phi^-|^R_S\,|\Phi^+|^r_s(c'(\cb),\cb)\,\delta^{0|8}(\etab_I+c_{I\,i}\etab_i)\,\prod'_{I,i}\,\delta(\Cb_{I i})~,\label{linkrep3}\ee
where on the right hand side, $c'$ have been solved in terms of $m$ remaining link variables, $\cb$, and the remaining integrations are
completely fixed by the sextics.

The two determinants contain $k{-}1$ powers of $\l\,,\,\r$ and $n{-}k{-}1$ powers of $[\,,\,]$, while in the overall kinematic denominator,
there are $(k{-}2)$ $[\,,\,]$'s and $(n{-}k{-}2)$ $\l\,,\,\r$'s, thus, dressed with the momentum-conservation delta functions, the amplitude is
indeed a pure number. The link representation, (\ref{linkrep1}),(\ref{linkrep2}) and (\ref{linkrep3}), is fully supersymmetric
as~(\ref{twistorstring}), and it also encodes other important symmetries and properties of gravity amplitudes, as we discuss now.

Under parity, we go to the representation $(\lambdab_a,\lambda_a,\eta_a):=(\lambdab_a,\Lambda_a)$ by exchanging $\lambda$ with $\lambdab$, and
Fourier transforming $\etab$'s to $\eta$'s. 
If we define conjugate link variables, $c^{\bot}_{i I}:=-c_{I i}$~\cite{Grassmannian}, and exchange the role of $N$ and $P$ (thus $k$ with
$n{-}k$, $I,J,R,S\in P$, $i,j,r,s\in N$) at the same time, we have the transformation \be \Phi^{\pm}\rightarrow (\Phi^{\mp})^\bot~,\quad
\delta^2(\lambda_i-c_{I i}\lambda_I)\,\delta^{2|8}(\Lambdab_I+c_{I i}\Lambdab_i)\rightarrow \delta^2(\lambdab_i+c_{i
I}^{\bot}\lambdab_I))\,\delta^{2|8}(\Lambda_I-c^{\bot}_{i I}\Lambda_i)~.\ee It is trivial to check that the integral measure, together with $J$
and $\delta(\Cb)$, is invariant under the transformation, thus the parity-conjugate of N$^{k{-}2}$MHV amplitudes, (\ref{linkrep2}), reads, \be
\MM^{\textrm{conj}}_{n,k}(\lambdab_a,\lambda_a,\eta_a)=\int d^{k\times(n-k)}
c^{\bot}\,J^{\bot}\,|(\Phi^+)^{\bot}|^R_S\,|({\Phi^-})^{\bot}|^r_s\,\delta^2(\lambdab_i+c^{\bot}_{i
I}\lambdab_I)\,\delta^{2|8}(\Lambda_I-c^{\bot}_{i I}\Lambda_i)\,\prod'_{i,I}\,\delta(\Cb^{\bot}_{i I})~,\ee which is exactly the formula for
N${}^{n{-}k{-}2}$MHV amplitudes, $\MM_{n,n{-}k}(\lambdab_a,\lambda_a,\eta_a)$, thus our formula is manifestly parity symmetric.

Moreover, the formula is clearly independent of the choice of $R,S$ or $r,s$, and, as manifested by the structure as product of two
determinants/trees, it is symmetric under permutations of $N$ and that of $P$, $S_k\times S_{n{-}k}$.~\footnote{(\ref{linkrep2}) are
particularly convenient for computing amplitudes of $k$ negative-helicity gravitons in $N$ and $n{-}k$ positive-helicity ones in $P$. By setting
$\eta_i=0$ and integrate over all $\eta_I$'s, the fermionic delta functions becomes trivial, and the $S_k\times S_{n{-}k}$ symmetry of graviton
scattering is manifest.} Of course, different choices of $N$ and $P$ give the same result, thus (\ref{linkrep1}) and (\ref{linkrep2}), are in
fact symmetric under permutations of all external particles. The $S_n$ symmetry can be made manifest by going back to a form using $n\times n$
matrices which depend on reference variables, similar to (\ref{Phi}), possibly at the expense of simplicity.

Last but not least, we can rewrite the link representation as a contour integral,\be \MM_{n,k}(\ZZ_I, \WW_i)=\oint_{\mathbf{\Cb}=0} d^{k\times
(n{-}k)} c_{I i}\,J |\Phi^-|^R_S\,|\Phi^+|^r_s\exp[ i\,c_{I i} \WW_i\cdot\ZZ_I]\prod'_{I,i}\frac 1 {\Cb_{I i}},\ee where the contour corresponds
to simultaneously set all $m$ sextics, $\Cb$'s, to zero. In this form, it is natural to consider a more general object, \be
\mathcal{G}_{n,k}(\ZZ_I, \WW_i)=\int d^{k\times (n{-}k)} c_{I i}\,J |\Phi^-|^R_S\,|\Phi^+|^r_s\exp[ i\,c_{I i} \WW_i\cdot\ZZ_I]\prod'_{I,i}\frac
1 {\Cb_{I i}}~\label{gnk},\ee which should be regarded as a multi-dimensional contour integral. Due to the $2\,n{-}4$ independent delta
functions as in~\ref{linkrep2}, the integral~(\ref{gnk}) is $m$-dimensional, and in principle we can evaluate its generic residues by setting
any $m$ poles to vanish.

In the simplest case, the MHV amplitude (or $\overline{\textrm{MHV}}$ amplitude which is related by parity), there is no integral needs to be
performed, and a direct evaluation reproduces immediately the well-known results in~\cite{Nguyen:2009jk}~\cite{Hodges:2012ym}. Beyond the MHV
case, while the tree amplitude can be computed by summing over residues corresponding to $\Cb_{I i}=0$, by the global residue theorem of
multi-dimensional contour integral~\cite{Grassmannian}, it can also be obtained using e.g. the contour which encloses all poles in~(\ref{gnk}),
except those poles $\Cb_{I i}=0$.

Although we have not computed such residues for $k>2$, from what we learnt in SYM case~\cite{Nandan:2009cc}~\cite{ArkaniHamed:2009dg}, we expect
that the computation is greatly simplified by summing over these residues, which gives different, but equivalent representations of tree
amplitudes, such as those obtained by BCFW recursion relations. It is natural to suspect that generic residues of~(\ref{gnk}) are associated
with leading singularities of loop-level amplitudes in $\mathcal{N}=8$ supergravity (see~\cite{Cachazo:2008dx}), and it is highly desirable to
compute and understand them. We leave the study of residues of~(\ref{gnk}), and their physical interpretations, to future works.

\section{Outlook: towards a Grassmannian formulation}

Having established a link representation for tree amplitudes in $\mathcal{N}=8$ supergravity, now we recast our formulae in a GL($k$) invariant
way, and we hope it will give us more hints on how to write down a Grassmannian contour integral which goes beyond tree amplitudes. As we have
learnt from the Grassmannian formulation for SYM amplitudes~\cite{Grassmannian}, we introduce a $G(k,n)$ Grassmannian, which, up to GL($k$)
transformations, can be described by $k\times n$ matrices $C_{\alpha a}$, with $\alpha=1,...,k$, $a=1,...,n$. A particular way for gauge-fixing
$C_{\alpha a}$ is to use a GL($k$) transformation to set the $k$ columns with $a=I \in N$ to be an identity matrix. Then the the remaining
non-trivial entries of $C_{\alpha a}$ can be identified with $c_{I i}$, and we call the entire gauge-fixed $k\times n$ matrix, including the
identity matrix part, $c_{\alpha a}$. This can be done by inserting an identity~\cite{ArkaniHamed:2009dg}, \be 1=\frac 1 {\vGL(k)}\int
d^{k\times n} C_{\alpha a}\int \frac{d^{k\times k} L_{\alpha}^{\beta}}{|L|^k} \,\prod_{\alpha,a}\delta(C_{\alpha a}-L^{\beta}_{\alpha}c_{\beta
a})~.\ee

By integrating out $c_{I i}$ and $L_{\alpha}^{\beta}$, the $k\times n$ delta function force us to set $L_{\alpha}^I=C_{\alpha I}$ for $I\in N$,
and set the link variables (also, as a consequence, its $2\times2$ determinants), to be written in terms of $k\times k$ minors of $C$, $(a_1 a_2
... a_k):=\epsilon^{\alpha_1...\alpha_k} C_{\alpha_1 a_1}C_{\alpha_2 a_2}...C_{\alpha_k a_k}$, \be c_{I i}=\frac{(\bar I\,i)}{(N)}~,\quad c^{I
J}_{i\,j}=\frac{(\overline{I J}\,i j)}{(N)}~\label{linktoC},\ee where $(N)=|L|$ is the minor with all $k$ columns in $N$ (in a canonical
ordering), $(\bar I\,i)$ means the $k\times k$ minor with column $I$ deleted and $i$ inserted, and similarly for $(\overline{I J}\,i j)$. The
sextics can also be written in a GL($k$)-invariant form, \be C_{R S I\,r s i}=\frac{S_{R S I\,r s i}}{(N)^4}~,\quad S_{I J K i j k
}=(\overline{I J} i j)(\overline{I K} j k)(\bar K i)(\bar J k)-(\overline{I J} j k)(\overline{I K} i j)(\bar K k)(\bar J i)~\label{sexttoC}.\ee
Note we pick up an Jacobian $|L|^{k{-}n}=(N)^{k{-}n}$ from the integration over $c$'s.

In addition, essentially identical to the SYM case, the delta functions in~(\ref{linkrep2}) can also be written in a GL$(k)$ invariant manner,
and it gives an Jacobian $(N)^{-4}$ due to the fact that the numbers of bosonic and fermionic delta functions are
unbalanced. 
Collecting everything (except a possible overall sign), we can put~(\ref{gnk}) in a GL($k$)-invariant form, \be \mathcal{G}_{n,k}=\frac 1
{\vGL(k)}\int d^{k\times n} C_{\alpha a}\frac{H|\Phi^-|^r_s|\Phi^+|^R_S}{S_1\,\ldots\,S_m}\,\int d^{2k}\rho \prod_a\delta^2(\rho_{\alpha}
C_{\alpha a}-\lambda_a)\prod_{\alpha} \delta^{2|8}(C_{\alpha a}\Lambdab_a)~\label{glk},\ee where in the denominator we list all $m$ sextics, and
we misuse the notation that $\Phi^{\pm}$ denote the following matrices, \ba \Phi^-_{I J, I\neq J}&=\displaystyle \frac{\l I\,J\r}{(\overline{I
J}\,r s)}~,\quad \Phi^-_{I I}&=-\sum_{J\neq I}\Phi^-_{I J}\,\frac{(\bar I\,r)\,(\bar I\,s)}{(\bar J\,r)\,(\bar J\,s)}~,\nl \Phi^+_{i j, i\neq
j}&=\displaystyle \frac{[ i\,j ]}{(\overline{R S}\,i j)}~,\quad \Phi^+_{i i}&=-\sum_{j\neq i}\Phi^+_{i j}\,\frac{(\bar R\, i)\,(\bar
S\,i)}{(\bar R\, j)\,(\bar S\,j)}~,\ea and finally by factoring out all $(N)$'s from (\ref{linktoC}) and (\ref{sexttoC}), we have an overall
factor \be H=\frac{(\overline{R S} r s)}{(N)^3(\bar R r)\,(\bar R s)\,(\bar S r)\,(\bar S s)}\,\prod'_{I i}\frac{(\overline{R S} r s)\,(\bar I
r)\,(\bar I s)\,(\bar R i)\,(\bar S i)}{(\bar I i)}~.\ee

There are some remaining steps to go from formula~(\ref{glk}) to a satisfying Grassmannian formulation. As we see from (\ref{linktoC}), the
formula~(\ref{glk}) only uses minors involving mostly columns in $N$, which, unlike the Grassmannian formula in SYM~\cite{Grassmannian}, does
not treat all minors of $C_{\alpha a}$ on an equal footing. Since the result is independent of the choice of $N,P$, we believe that~(\ref{glk})
can be put into a manifestly $S_n$-invariant form. In addition, as is well known~\cite{Nandan:2009cc}~\cite{Dolan:2010xv}, there are various
relations satisfied by the sextics~(\ref{sexttoC}), e.g. for $k=3$,\be \delta(S_{a b c d e f})\,\delta(S_{a b c d e g})=\frac{(b c f)\,(a d
f)}{(b c e)\,(a d e)}\,\delta(S_{a b c d e f})\,\delta(S_{a b c d f g})~.\ee Thus, by rewriting the product of $m$ sextices using such
relations, we can derive equivalent GL($k$)-invariant formulae of tree amplitudes, which enjoy certain nice properties, e.g. a manifest particle
interpretation~\cite{ArkaniHamed:2009dg}. Eventually, from~(\ref{glk}), we hope to obtain a Grassmannian contour integral which contains
beyond-tree-level information, and manifests important symmetries and properties of gravity amplitudes.

\acknowledgments It is a pleasure to thank F.~Cachazo, B.~Feng, A.~Hodges, L.~Mason and D.~Skinner for very interesting discussions on various
related subjects. While completing the project, I was informed by F.~Cachazo, L.~Mason and D.~Skinner of their work in progress, which has some
overlap with the present note.

\end{document}